\documentclass[pre,floatfix,superscriptaddress,twocolumn,9pt]{revtex4-1}
    
\usepackage{amsmath}
\usepackage{amssymb}
\usepackage{amsfonts}
\usepackage{graphicx}
\usepackage{color}
\usepackage{bm}
\usepackage[english]{babel}

\renewcommand{\epsilon}{\varepsilon}

\newcommand{\figurewidth}{.47\textwidth}

\newcommand{\kb}{\ensuremath{k_{\mathrm{B}}}}
\newcommand{\kT}{\ensuremath{\kb T}}
\newcommand{\del}{\ensuremath{\nabla}}

\newcommand{\ww}{\mathbf{w}}

\newcommand{\thetitle}{Learning Free Energy Landscapes Using Artificial Neural Networks}

\begin{document}

\author{Hythem Sidky}
\affiliation{%
  Department of Chemical and Biomolecular Engineering, %
  University of Notre Dame, %
  Notre Dame, IN 46556
}
\author{Jonathan K. Whitmer}
\email{jwhitme1@nd.edu}
\affiliation{%
  Department of Chemical and Biomolecular Engineering, %
  University of Notre Dame, %
  Notre Dame, IN 46556
}

\title{\thetitle}

\begin{abstract}
  Existing adaptive bias techniques, which seek to estimate free
  energies and physical properties from molecular simulations, are
  limited by their reliance on fixed kernels or basis sets which 
  hinder their ability to efficiently conform to varied free energy 
  landscapes.  Further, user-specified parameters are in general 
  non-intuitive, yet significantly affect the convergence rate and 
  accuracy of the free energy estimate.  Here we propose a novel method 
  wherein artificial neural networks (ANNs) are used to develop an 
  adaptive biasing potential which learns free energy landscapes. 
  We demonstrate that this method is capable of rapidly adapting to 
  complex free energy landscapes and is not prone to boundary or oscillation 
  problems. The method is made robust to hyperparameters and overfitting 
  through Bayesian regularization which penalizes network weights and 
  auto-regulates the number of effective parameters in the network.  
  ANN sampling represents a promising innovative approach which can 
  resolve complex free energy landscapes in less time than conventional 
  approaches while requiring minimal user input.
\end{abstract}

\maketitle

\section{Introduction}

Use of adaptive biasing methods to estimate free energies of
relevant chemical processes from molecular simulation has gained
popularity in recent years. Free energy calculations have become 
central to the computational study of biomolecular 
systems~\cite{Barducci03122013}, material properties~\cite{Joshi2014}, 
and rare events~\cite{Valsson2016}. The metadynamics
algorithm~\cite{Laio2002}, which uses a sum of Gaussian kernels to
obtain a free energy landscape (FEL) on-the-fly, has unleashed a groundswell of
new activity studying on--the--fly approximations of FELs. Many extensions
of this method have been proposed~\cite{Barducci2008, Singh2011,
Dama2014, Valsson2016}, which enhance the speed and accuracy of
metadynamics simulations, though the essential
limitation---representing the FEL with Gaussians---has
remained. Notably, the optimal choice of parameters such as the
multidimensional Gaussian widths, height and deposition rate are
unknown for each new system and poor choices can severely affect
both the convergence rate and the accuracy of the estimated
FEL. Partial remedies involving locally optimized bias
kernels exist~\cite{Branduardi2012, Barducci2008}, though these cannot
correct for systematic limitations inherent to the approximation of a
surface by Gaussians, including boundary errors~\cite{McGovern2013}.

Inspired by adaptive bias umbrella sampling~\cite{Mezei1987, Hooft1992, Bartels1997, Bartels1998}, 
algorithms have begun to emerge which address these limitations, including variationally 
enhanced sampling (VES)~\cite{Valsson2014}, basis function sampling
(BFS)~\cite{Whitmer2014}, and Green's function sampling
(GFS)~\cite{Whitmer2015}. Each utilizes orthogonal basis
expansions to represent the underlying FEL, with VES deriving a
variational functional which, when stochastically minimized, yields
the optimal expansion coefficients. The related BFS approximates FELs
using a probability density obtained through an accumulated biased
histogram, and determines the coefficients by direct projection of the
data onto a chosen orthogonal basis set. GFS, a dynamic version of
BFS, invokes a metadynamics-like algorithm approximating a delta
function, resulting in a trajectory which reveals optimal coefficients
on-the-fly. The basis expansions employed by these methods, which can
represent the often nuanced and highly nonlinear FELs more robustly
than Gaussians, also benefit by eliminating the need for
kernel-specific properties such as width and height. A user must still
specify the number of terms to include in the expansion and 
``deposition rate'' parameters which control the aggressiveness of bias updating. 
Still, there remain a number of non--ideal practical and theoretical issues which 
must be addressed. Importantly, the use of \emph{any} functional form to represent 
the FEL inherits both the advantages and disadvantages of that set of functions. 
Basis sets, for example, may lead to Runge or Gibbs phenomena on sharply varying surfaces or
near boundaries. These characteristic oscillations introduce artificial 
free energy barriers which may prevent the system from sampling important regions 
of phase space, and result in a non-converging bias. These do not self-correct, and
increasing the order of the expansion is generally not helpful~\cite{Kincaid2009}. 

Alternative algorithms, while less prevalent in the literature,
nonetheless exhibit many benefits. Adaptive biasing force (ABF)
methods~\cite{Darve2001}, developed in parallel to metadynamics-derived
methods, seek to estimate the underlying mean force (rather than
the free energy) at discrete points along the coordinates of
interest. The mean force is then integrated at the end of a simulation
to obtain the corresponding FEL. The primary advantage of ABF over
biasing potential approaches is the local nature of the mean force at
a given point. While the free energy is a global property related the
to probability distribution along a set of coordinates, the mean force
at a given point is independent of other regions. As a consequence,
ABF methods have proven to be formidable alternatives to biasing
potentials and are recognized to rapidly generate accurate FEL
estimates~\cite{Comer2015}. Though powerful in practice, these have
seen limited application due to the complexity of the original
algorithm~\cite{Darve2001} and restrictions on the type of coordinates
which may be sampled~\cite{Comer2015, Fu2016}.

Recognizing the distinct strengths and weaknesses of existing algorithms, we present an
adaptive biasing potential algorithm also utilizing artificial neural networks
(ANNs) to construct the FEL. ANNs are a form of \emph{supervised learning},
consisting of a collection of activation functions, or neurons, which combine
to produce a function that well--aproximates a given set of
data~\cite{Demuth2014}.  The use of ANNs in the form of ``deep learning'' has
permeated virtually every quantitative discipline ranging from medical
imaging~\cite{gletsos2003} to market forecasting~\cite{pao2007}. ANNs have
also been applied in various ways in molecular simulation to fit \emph{ab
  initio} potential energy surfaces~\cite{Behler2007, Khaliullin2010},
identify reaction coordinates~\cite{Ma2005}, obtain structure-property
relationships~\cite{So1996}, and determine quantum mechanical forces in
molecular dynamics (MD) simulations~\cite{Li2015}. Recently, advances in
machine learning applied to molecular simulations have aided in the
reconstruction of FELs. As a competitive alternative to the weighted histogram
analysis method (WHAM)~\cite{Kumar1992}, Gaussian process regression was used
to reconstruct free energies from umbrella sampling
simulations~\cite{Stecher2014}. Adaptive Gaussian-mixture umbrella sampling
was also proposed as a way to identify free energy basins in complex
systems. Similar to metadynamics, this approach relies on Gaussian
representations of the free energy surface, with a focus on resolving low
lying regions in higher dimensional free energy space.  Very
recently~\cite{Galvelis2017}, ANNs were combined with a nearest--neighbor
density estimator for high dimensional free energy estimation. They were also
used to post-process free energy estimates obtained from enhanced sampling
calculations, providing a compact representation of the data for
interpolation, manipulation and storage.~\cite{Schneider2017} To the best of
our knowledge, the only use of ANNs to estimate free energies on-the-fly was
the very recent work by Galvelis and Sugita~\cite{Galvelis2017}. Although our
proposed method also utilizes ANNs, it differs critically in their
application.  The Bayesian NN-based free-energy method reported here rapidly
adapts to complex FELs without spurious boundary and ringing artifacts,
retains all accumulated statistics throughout a simulation, is robust to the
few user-specified hyperparameters and flexible enough to predict FELs where
key features may be unknown before simulation.

\section{Methods}

\subsection{Artificial neural networks}

Inspired by biological neurons, artificial neural networks (ANNs) are a 
collection of interconnected nodes which act as universal approximators;
that is, ANNs can approximate any continuous function to an arbitrary 
level of accuracy with a finite number of neurons.\cite{Hornik1989} 
ANNs are defined by a network topology which specifies the number of neurons 
and their connectivity. Learning rules determine how the weights
associated with each node are updated. Layers within a neural network
consist of sets of nodes which receive multiple inputs from the
previous layer and pass outputs to the next layer. Here we utilize a 
fully connected network, where every output within a layer is an input 
for every neuron in the next layer. 

Mathematically, the activation $F_i$ of a fully--connected layer $k$
can be phrased as 
\begin{equation}
  F_i^k = S\left(\sum_{j=1}^{M}{w_{ji}^k F_j^{k-1}} + b_i^k\right),
\end{equation}
\noindent where $w_{ji}$ and $b_i$ terms define the layer weights and biases
respectively. Indices $i$ and $j$ represent the number of neurons in
the current and previous layers respectively. The activation function
$S(x)$ is applied element-wise and is taken to be $S(x) = \tanh(x)$, with the 
exception of the output layer which utilizes the linear activation $S(x) = x$.
Backpropagation\cite{MacKay1992} is used to update the network weights, and to 
obtain the output gradient with respect to the inputs for force calculations within 
molecular dynamics simulations.

In this work, the network seeks the free energy of a set of collective variables 
(CVs). We define a CV ($\xi$) via the mapping, $\xi : \mathbb{R}^{3N} \rightarrow
\mathbb{R}$ from the $3N$ positional coordinates of a molecular system
onto a continuous subset of the real numbers. This can be any property
capturing essential behavior within the system, such as intermolecular distances, 
energies, or deviation from native protein structures. The set of collective 
variable coordinates and corresponding approximate free energies $\{ \{ \bm{\xi}_1, \tilde{F}_1\}, 
\{\bm{\xi}_2, \tilde{F}_2\}, \ldots, \{\bm{\xi}_N, \tilde{F}_N\}\}$ represent the 
training data, and obtaining them is a key component of this method described in 
Section~\ref{sec:sampling}. 

Training a neural network involves finding the optimal set of weights 
and biases that best represents existing data while maximizing generalizability 
to make accurate predictions about new observations. In particular, one seeks 
to find the least complex ANN architecture that can capture the true 
functional dependence of the data. However, since it is nearly impossible to 
know the proper architecture for an arbitrary problem beforehand, several 
strategies exist to ensure that a sufficiently complex neural network does 
not overfit training data. One common approach known as \emph{early stopping} 
involves partitioning data into training and validation sets. Network optimization 
is performed using the training data alone, terminating when the validation error 
begins to increase. Another approach, referred to as regularization or weight decay, 
penalizes the $L_2$ norm of the network weights which is analogous to reducing the 
number of neurons in a network.\cite{Demuth2014} In practice, both of these strategies 
are often used simultaneously.

Although early stopping and regularization are great approaches for post-processing 
and offline applications, they are not well-suited for on-the-fly sampling. Ideally 
one would like to utilize all of the statistics collected over the course of a simulation 
in constructing the free energy estimate without needing to subsample the data for 
validation. More importantly, the optimal choice of network architecture, regularization 
strength, and other hyperparameters will vary from system to system and typically require 
the use of cross-validation to be identified. Hyperparameter optimization will not only 
make a free energy method prohibitively expensive, it will also require significant 
user input, negating many of the advantages offered over existing enhanced sampling methods. 
Other choices such as training loss function and optimization algorithm can affect the 
quality of fit and speed of convergence. In what follows, we show how an appropriate 
choice of loss function and optimization algorithm combined with a Bayesian framework 
yields a self-regularizing ANN which uses all observations for training, is robust to 
hyperparameter choice, maintains generalizability, and is thus ideal for use in an 
adaptive sampling method. 

\subsection{Bayesian regularization}

We begin by defining the loss function with respect to the network weights $\ww$
for a set of observations $\{\bm{\xi}_i, \tilde{F}_i\}$ which are the CV coordinates 
and free energy estimates, 

\begin{equation}
E(\ww) = \beta E_D + \alpha E_W = 
\beta \sum_{i=1}^{N}{(\tilde{F}_i - \hat{F}_i)^2}  + \alpha \sum_{i=1}^{K}{w_i^2}.
\end{equation}

\noindent The index $i$ in this equation runs over all $N$ bins (in the first
term) and all $K$ weights and biases within the network (second term). The loss function
can be decomposed into two terms: a sum squared error $E_D$  between the
network predictions $\hat{F}_i$ and targets $\tilde{F}_i$, and a  penalty or
regularization term $E_W$ on the network weights and biases $w_i$.  The ratio
$\alpha/\beta$ controls the complexity of the network, where an increase
results in a smoother network response.

Following the practical Bayesian framework by MacKay~\cite{MacKay1992, MacKay1992a}, 
we assume that the free energy estimates, which are the target outputs, are generated by

\begin{equation}
	\tilde{F}_{i} = F(\mathbf{x}_i) + \epsilon_i, 
\end{equation}

\noindent where $F(\cdot)$ is the true (as of yet unknown) underlying free energy surface 
and $\epsilon_i$ is random, independent and zero mean sampling noise. The training 
objective is to approximate $F(\cdot)$ while ignoring the noise. The Bayesian 
framework starts by assuming the network weights are random variables, and we 
seek the set of weights that maximize the conditional probability of the weights 
given the data $D$. It is useful, and perhaps more intuitive, to think about 
the distribution of weights as a distribution of possible functions represented 
by a given neural network architecture, $A$. Invoking Bayes' rule yields:

\begin{equation} 
	P(\ww | D, \alpha, \beta, A) = 
	\frac{P(D|\ww,\beta, A)P(\ww|\alpha, A)}{P(D|\alpha, \beta, A)}.
	\label{eq:posterior1}
\end{equation}

Although our notation resembles that of the Boltzmann distribution, its use here is purely statistical.
Since we assumed independent Gaussian noise on the data, the probability density, which is our 
likelihood, becomes 
\begin{equation}
	P(D|\ww, \beta, A) = \frac{1}{Z_D(\beta)}\exp(-\beta E_D), 
	\label{eq:prior1}
\end{equation}
\noindent where $\beta = 1/(2\sigma_\epsilon^2)$, $\sigma_\epsilon^2$ is the variance of 
each element of $\epsilon_i$, and $Z_D(\beta) = (2 \pi \sigma_\epsilon^2)^{N/2} = (\pi/\beta)^{N/2}$.
The prior density $P(\ww|\alpha, A)$ of the network weights is assumed to be a zero mean 
Gaussian, 

\begin{equation}
	P(\ww | \alpha, A) = \frac{1}{Z_W(\alpha)}\exp(-\alpha E_W),
\end{equation}

\noindent with $\alpha = 1/(2 \sigma_w^2)$ and $Z_W = (\pi / \alpha)^{K/2}$. We can 
rewrite the posterior density (Eq.~\ref{eq:posterior1}) as 

\begin{equation}
	P(\ww | D, \alpha, \beta, A) =  
	\frac{1}{Z_W(\alpha)} \frac{1}{Z_D(\beta)} \frac{\exp(-(\beta E_D + \alpha E_W))}{P(D|\alpha, \beta, A)}.
	\label{eq:posterior2}
\end{equation}

The evidence $P(D|\alpha, \beta, A)$ is not dependent on the weights and we can combine 
the normalization constants to obtain

\begin{equation}
	P(\ww | D, \alpha, \beta, A) = \frac{1}{Z_F(\alpha, \beta)} \exp(-E(\ww)).
	\label{eq:posterior3}
\end{equation}

\noindent It is now clear that maximizing the posterior density with the appropriate 
priors is equivalent to minimizing our loss. $\alpha$ and $\beta$ also take on a 
statistical meaning. $\beta$ is inversely proportional to the variance of the sampling 
noise. If the variance is large, $\beta$ will be small and force the network weights 
to be small resulting in a smoother function, attenuating the noise. 

What remains is to estimate $\alpha$ and $\beta$ from the data: 
\begin{equation}
	P(\alpha, \beta | D, A) = 
	\frac{P(D|\alpha, \beta, A)P(\alpha, \beta | A)}{P(D | A)}.
\end{equation}
\noindent Assuming a uniform prior density on $\alpha, \beta$ and using 
Equations~\ref{eq:posterior2} and~\ref{eq:posterior3} it follows that
\begin{equation}
	P(D|\alpha, \beta, A) = \frac{Z_F(\alpha, \beta)}{Z_W(\alpha) Z_D(\beta)}.
	\label{eq:combinedprobs}
\end{equation}
\noindent The only remaining unknown is $Z_F(\alpha, \beta)$ which can be approximated 
via a Taylor expansion of the loss function about the maximum probability weights $\ww^{MP}$,
\begin{equation}
	E(\ww) \approx E(\ww^{MP}) + \frac{1}{2}(\ww - \ww^{MP})^T \mathbf{H}^{MP}(\ww - \ww^{MP}),
\end{equation}
\noindent where $\mathbf{H} = \beta \del^2 E_D + \alpha \del^2 E_W$ is the Hessian. 
Substituting the expansion back in Equation~\ref{eq:posterior3} gives, 
\begin{widetext}
\begin{equation}
    P(\ww | D, \alpha, \beta, A) \approx \frac{1}{Z_f(\alpha, \beta)} 
	\exp(-E(\ww^{MP}) -  \frac{1}{2}(\ww - \ww^{MP})^T \mathbf{H}^{MP}(\ww - \ww^{MP})).
\end{equation} 
\end{widetext}
\noindent Equating this expression to the standard form of a Gaussian density yields 
$Z_f(\alpha, \beta) \approx (2 \pi)^{K/2} \det(\mathbf{H^{MP}}^{-1})^{1/2} \exp(-E(\ww^{MP}))$.
Finally, setting the derivative of the logarithm of Equation~\ref{eq:combinedprobs} to zero 
and solving for optimal $\alpha, \beta$ gives, 
\begin{align}
	\alpha^{MP} &= \frac{\gamma}{2 E_W(\ww^{MP})}  &  \beta^{MP} & = \frac{N-\gamma}{2 E_D(\ww^{MP})},
\end{align}
\noindent with $\gamma = K - 2 \alpha^{MP} tr(\mathbf{H}^{-1})$, which also represents the 
effective number of parameters in the network used to reduce the error function. Starting 
with $\gamma = K$, the hyperparamters $\alpha, \beta, \gamma$ are updated iteratively within a 
Levenberg-Marquardt optimization routine~\cite{nocedal} where the Hessian is approximated 
from the error Jacobian $\mathbf{J}$ as $\mathbf{H} \approx \mathbf{J}^T \mathbf{J}$. 
Training proceeds until the error gradient diminishes, or the trust region radius 
exceeds $10^{10}$.

\subsection{Sampling method}

\label{sec:sampling}

\begin{figure}
  \begin{center}
    \includegraphics[width=\figurewidth]{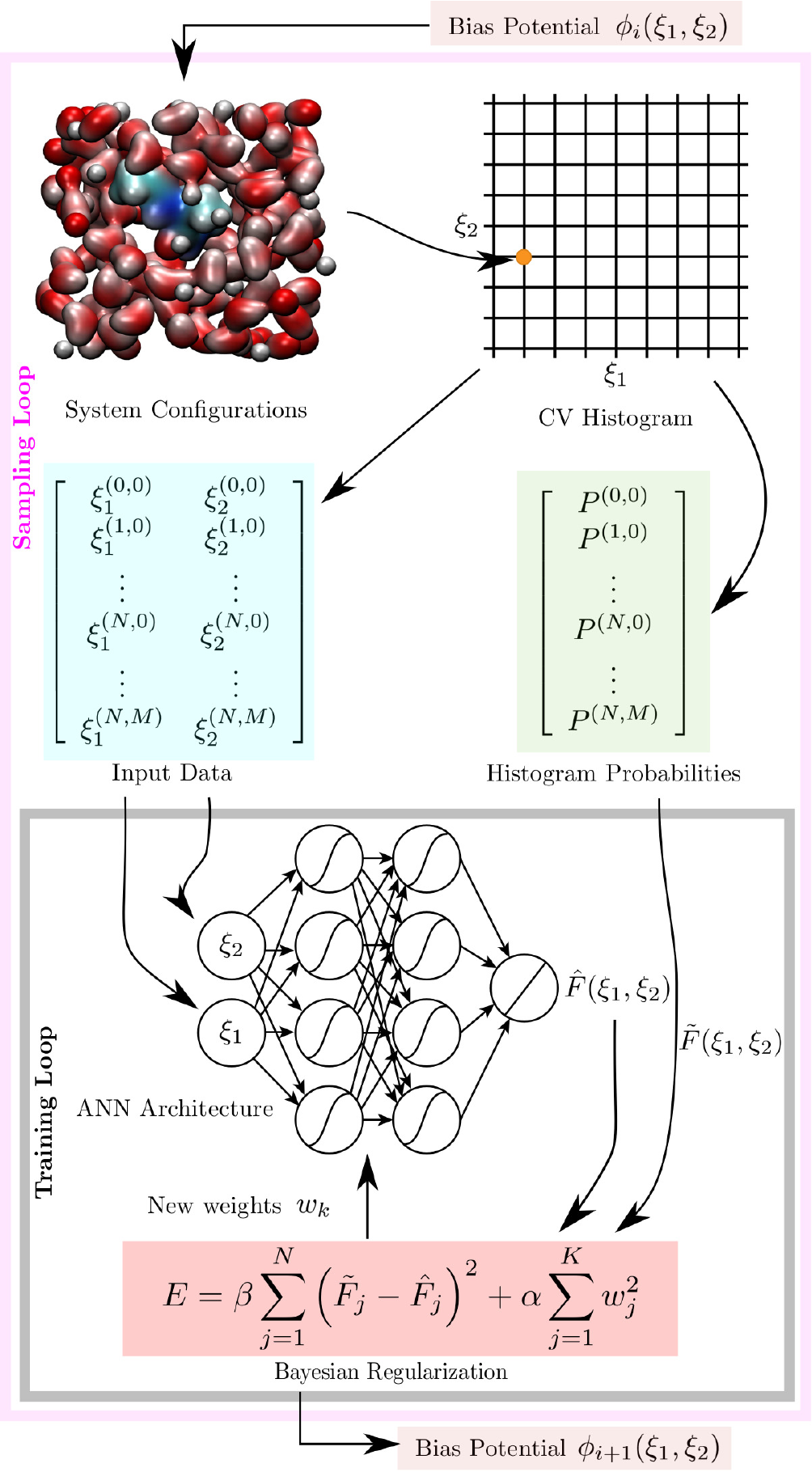}
  \end{center}
  \caption{Schematic of the ANN adaptive biasing method. Hits along the
    collective variable(s) are recorded onto a histogram during the
    simulation.  The discrete grid coordinates and corresponding reweighted
    normalized probabilities are used as the input and output training data
    respectively.  An ANN architecture is chosen prior to runtime, and is
    trained iteratively using a Bayesian regularization procedure. The
    objective function chosen represents the sum squared error between the
    free energy estimate $\tilde{F}$ and ANN output $\hat{F}$. 
    Hyperparameters $\alpha, \beta$ represent the relative penalties to the data and 
    network weights respectively which are automatically adjusted during the 
    optimization routine.}
  \label{fig:annpanel}
\end{figure}

Figure~\ref{fig:annpanel} shows a schematic of the ANN sampling method.
For a given CV $\xi$, the free energy along the CV is defined as
\begin{equation}
F(\xi) = -\kT \log{P(\xi)}.
\end{equation} 
\noindent Here $\kb$ is Boltzmann's constant, $T$ is temperature, and
$P(\xi)$ is the probability distribution of the system along
$\xi$, which the ANN method seeks to learn. Simulations proceed in stages or sweeps,
indexed by $i$, where a bias, $\phi_i(\xi)$, is iteratively refined until it
matches $-F(\xi)$. Typically, a
simulation begins unbiased, though pre-trained networks may be loaded
if available. The value of $\xi$ is recorded at each step
into a sparse or dense histogram $H_i(\xi)$, depending on the CV dimensionality. 
Upon completion of a fixed number of steps, $N_\mathrm{s}$, this histogram is 
reweighted~\cite{Torrie1977} according the bias applied during that sweep,
\begin{equation}
\tilde{H}_i(\xi) = H_i(\xi)e^{\frac{\phi_i(\xi)}{\kT}},
\end{equation}
and subsequently added to a global partition function 
estimate, 
\begin{equation}
Z_i(\xi) = \sum_{j \leq i} \tilde{H}_j(\xi).
\label{eq:partition}
\end{equation}
\noindent Suitably scaled, this provides an estimator $\tilde{P}_i(\xi)$ with
associated free energy
\begin{equation}
  \tilde{F}_i(\xi)=-\kT\log\left[\tilde{P}_i(\xi)\right]
\end{equation}
approximating the true FEL $F(\xi)$. As illustrated, the $\tilde{F}_i(\xi)$
serve as a training set for the ANN, using the discrete CV coordinates as
input data. As a result, there are $N = \prod_{i=1}^{N_{CV}}M_i$ observations
where $M_i$ is the number of bins along each CV $i$, and $N_{CV}$ is the total
number of CVs.  Henceforth, we will use a compact notation to denote the ANN
architecture, with $\{5,2\}$ indicating a neural network with two hidden
layers containing 5 and 2 neurons respectively. The output layer contains a
single neuron defining a function $\hat{F}_i(\xi)$ comprising a best
representation of the estimated free energy from the known data.

The bias for the initial sweep, $\phi_0(\xi)$ is defined at the
outset. Subsequent sweeps utilize the ANN output via the equation
$\phi_{i+1}(\xi)=-\hat{F}(\xi)$, where $\hat{F}$ denotes the optimized output of 
the neural network. By construction, this biasing function is continuous, differentiable, and
free of ringing artifacts. Furthermore, one particular strength of this 
method is the capacity for a well--trained ANN to interpolate over the learned
surface; each successive $\phi_i(\xi)$ yields a good
estimate of the free energy even in poorly sampled regions,
necessitating fewer samples in each histogram bin to obtain a high
fidelity approximation of the FEL, as will be demonstrated below. The
entire process is then iterated until the output layer
$\hat{F}(\xi)$ achieves a measure of convergence to the true FEL $F(\xi)$. 

\section{Results}

\subsection{One dimensional FEL}
\begin{figure}
  \begin{center}
    \includegraphics[width=\figurewidth]{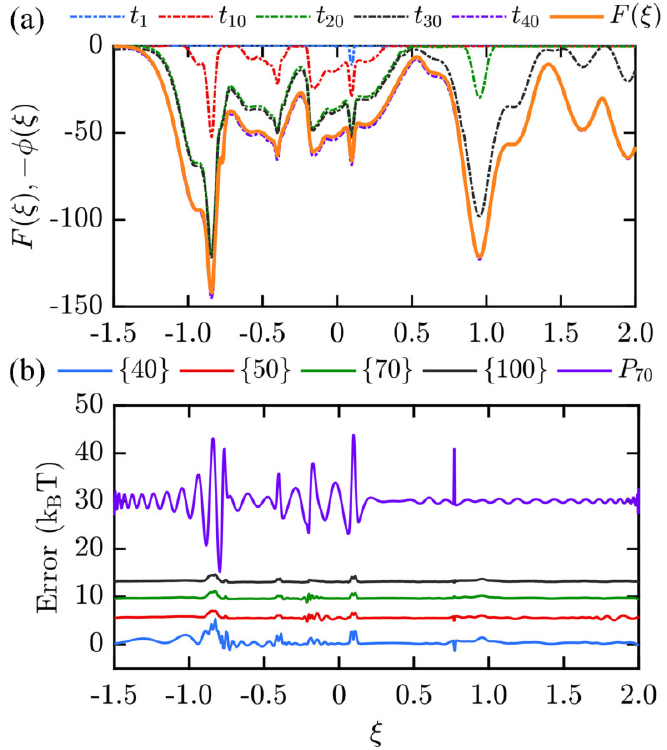}
  \end{center}
  \caption{ (a) Evolution of negative bias over time using a $\{40\}$
    network on a benchmark rugged free energy landscape. The resolved
    bias converges to the $-F(\xi)$ smoothly over time with no
    apparent edge effects or oscillations in regions of steep
    derivatives. An analysis of the bias error (b) for different network 
    sizes shows very good representation. There is a
    significant drop in error between $40$ and $50$ neurons after
    which the error signature stabilizes as a consequence of
    regularization. For comparison is a projection of the true surface 
    onto an order $70$ Legendre polynomial which represents an ideal
    result for methods employing polynomial basis sets. Values have
    been shifted along the y--axis for clarity.}
  \label{fig:mc1dpoc}
\end{figure}

We first illustrate the robustness of the ANN method by studying a
synthetic one dimensional Monte Carlo FEL composed of 50
Gaussians. The rugged landscape is meant to represent that of a glassy
system or protein folding funnel; this exact surface was utilized as a
benchmark for GFS~\cite{Whitmer2015}.  Figure~\ref{fig:mc1dpoc}(a) shows
the evolution of the ANN bias as a function of MC sweep $t$, where a
sweep consists of $10^5$ moves. Utilizing a $\{40\}$ network, the
analytical result is matched impressively within 40 sweeps ($4\times10^6$
total MC moves). This highlights how even a parsimonious ANN is
capable of capturing rugged topographies.

Further investigation into the effect of network size on the accuracy
of the bias yields interesting results. Figure~\ref{fig:mc1dpoc}(b) shows
the difference in $\phi(\xi)$ and $F(\xi)$ for network sizes from
$\{40\}$ to $\{100\}$. While the smallest network does not exceed $5
k_{\mathrm{B}}T$ in error, significant improvement is evident upon
increasing the network size to $\{50\}$. Beyond this, error does not
decrease significantly upon further refinement, indicating that
Bayesian regularization converges to an effective number of parameters, 
demonstrating good generalization and lack of overfitting. Also plotted is a 
projection of the FEL onto a $70$ degree Legendre polynomial, which represents 
an idealized result of basis expansion methods~\cite{Whitmer2014, Valsson2014,
Whitmer2015}. There are clear minor oscillations across the entire
interval and more pronounced ones in regions of sharp
derivatives. Edge effects are also present with a substantial amount
of error introduced at the right boundary. Remarkably, the ANN is capable 
of yielding a more accurate result using many fewer functions, while 
presenting no appreciable ringing. 

Many free energy landscapes for meaningful systems are considerably smoother than 
this one-dimensional example. However, since adaptive sampling methods are iterative, 
initial estimates of even smooth surfaces are often noisy and sharply varying. This 
presents a challenge for basis expansion methods as one must use a large enough
basis to drive the system out of metastable states yet one small enough that early sampling 
noise is effectively smoothed by low frequency truncation. Practically, the FEL
can be recovered from a reweighted histogram which affords some flexibility in basis
size, but the issue remains that enough terms must be retained to adapt to any features 
of significant size. The resulting emergence of oscillation artifacts greater than a few 
$k_\mathrm{B} T$ can unfortunately introduce significant sampling issues. 

\begin{figure*}[t!]
  \begin{center}
    \includegraphics[width=\textwidth]{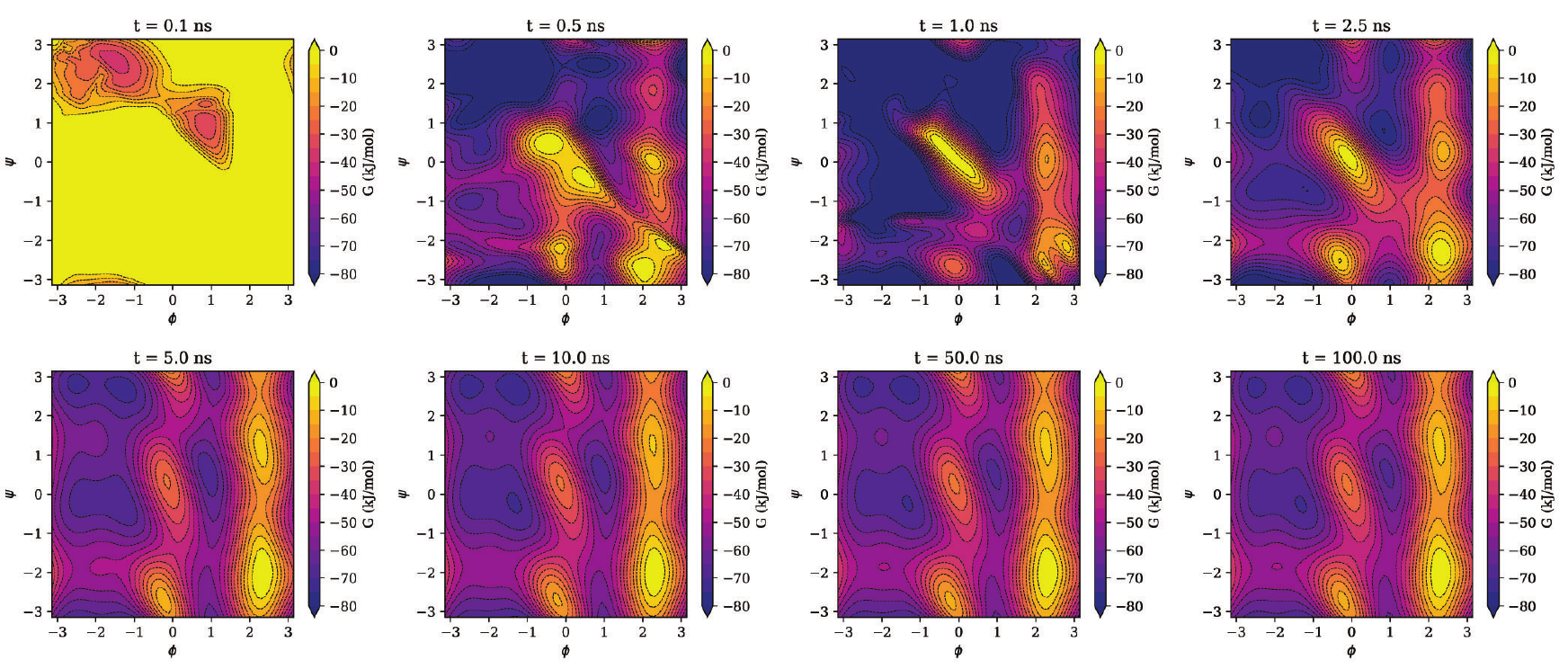}
  \end{center}
  \caption{
    Snapshots of the estimated free energy landscape as a function of
    the Ramachandran angles for alanine dipeptide in water at $298.15$~K
     using ANN sampling on a $60 \times 60$ grid. ANN sampling rapidly 
    resolves the free energy landscape. Much of the speed of ANN sampling 
    derives from its ability to smooth out statistical noise using the 
    Bayesian framework and interpolate well in under--sampled regions at 
    early times. Distortions are rapidly refined as the bias drives sampling
    to those regions and a stable approximation is achieved at approximately 
    $5\ \mathrm{ns}$, where it is only subject to statistical fluctuations.
    }
  \label{fig:annadp_ref}
\end{figure*}

\subsection{Two dimensional FEL: Alanine dipeptide in water}
As a more realistic example, we study the isomerization of alanine dipeptide
(ADP) in water. This system is a good testbed for free energy methods, as it
has a rich FEL with significant variation and has been thoroughly studied. It is 
important to note that the dynamics of ADP in water are \emph{significantly} slower 
than in vacuum, and is reflected in the convergence times of the free energy surfaces. 
The two torsional angles $\phi$ and $\psi$ of the ADP molecule are the CVs, which 
approximate~\cite{McCarty2017, Ma2005}, but are not precisely, the slow modes of the 
isomerization process. ADP simulations are carried out in the NPT ensemble with a 
timestep of 2 fs and hydrogen bond constraints using the LINCS algorithm. Long-range 
electrostatics are calculated with particle mesh Ewald on a grid spaced at 0.16 nm. 
A stochastic velocity rescaling thermostat is used with a time constant of 0.6 ps to 
maintain 298.15~K and a Parrinello-Rahman barostat with a time constant of 10 ps 
maintains a pressure of 1 bar. The system consists of a single ADP molecule 
and 878 water molecules described by the Amber99sb and TIP3P forcefields respectively. 
A modified version of SSAGES v0.7.5 compiled with Gromacs 2016.3 was used for all 
simulations.

A periodic grid containing $60 \times 60$ points is used with the ANN having topology 
$\{10,6\}$ and a chosen sweep interval of $10$ ps. Figure~\ref{fig:annadp_ref} 
shows snapshots of the estimated FEL at various time intervals. At early times one can see 
how the ANN is able to interpolate poorly--sampled and entirely unsampled areas within 
the vicinity of the initial exploration. The ANN bias is quickly refined as data
are collected in these regions. Statistical noise is smoothed via the Bayesian 
regularization which forces a smooth network response at early times by increasing 
weight decay. After only $2.5\ \mathrm{ns}$ the bias closely resembles the final FEL, 
with only minor distortion of features present at regions of low probability. 
It is clear that even with limited and noisy data the ANN is able to construct an 
idealized, smooth interpolation of the FEL. In fact, the unbiased histogram used to 
train the ANN remains noisy for a significant amount of time after the network
has converged. This is in stark contrast to other methods which rely on the unbiased 
histogram to recover an accurate FEL where the bias potential is unable to capture 
the fine details of the surface.

\begin{figure}
  \begin{center}
    \includegraphics[width=\figurewidth]{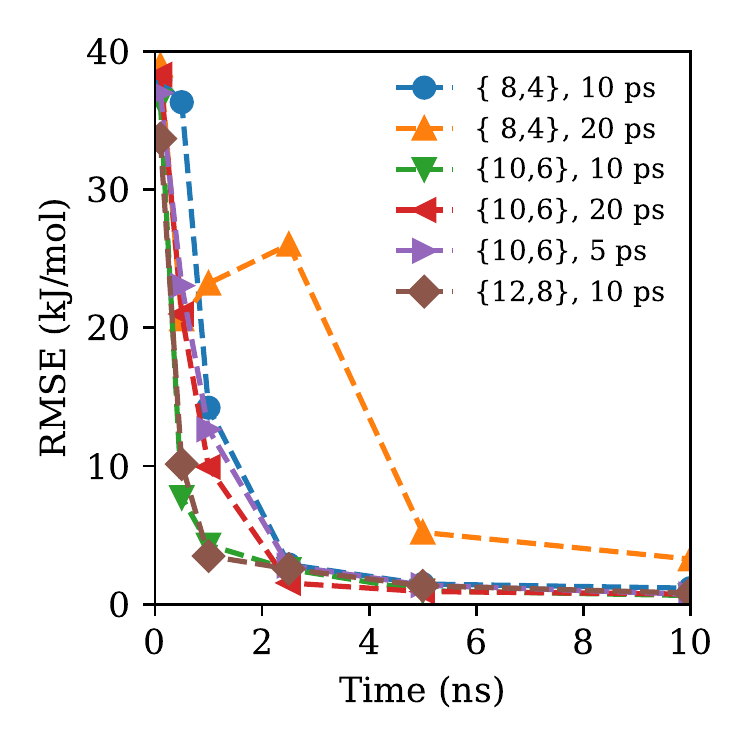}
  \end{center}
  \caption{
    Free energy landscape root mean squared error (RMSE) for alanine dipeptide 
    using various ANN architectures and sweep lengths as a function of time.
    The RMSE is computed relative to a reference FEL obtained from a $100$ ns 
    simulation. With the exception of the small $\{8, 4\}$ network with a large 
    sweep, all other systems quickly converge to within $1 k_\mathrm{B}T$ at
    $2.5$ ns, and within $1$ kJ/mol at $5$ ns. This demonstrates that beyond a 
    minimum network size and sweep, the performance of ANN sampling is insensitive 
    to a broad change of user-specified parameters.
    }
  \label{fig:annadp_rmsd}
\end{figure}

We investigate the sensitivity of ANN sampling to hyperparameters by studying 
six additional combinations of network architectures and sweep intervals.
The reference FEL is the final ANN state after $100$ ns of simulation from the previous example. 
The new systems were run on a $30 \times 30$ histogram, unlike the reference, to 
examine the behavior of ANN sampling with less available data and subject to larger 
discretization. Figure~\ref{fig:annadp_rmsd} shows the root mean square error (RMSE) of 
the FELs over time for the various systems. Except for a single setup, all FELs converged 
to within a $k_\mathrm{B} T$ at about $2.5$ ns, and $1$ kJ/mol at $5$ ns. The remaining 
configuration eventually converges at approximately $12$ ns. This relatively slow convergence 
represents the limit of network size and sweep length; a $\{8,4\}$ network is \emph{just}
able to represent the FEL, but requires more frequent updates and training. For the larger 
$\{10,6\}$ and $\{12,8\}$ networks, the sweep length has minimal impact on performance. 
For the end user, any reasonable choice of sweep length and a network of sufficient size 
should provide near optimal performance for new systems, eliminating the need for pilot 
simulations and significant prior experience. 

\subsection{Three dimensional FEL: Rouse modes of a Gaussian chain}

We compute the three--dimensional dense free energy landscape of the three 
longest wavelength Rouse modes~\cite{RubinsteinColby} for a 21 bead Gaussian 
chain. This unconventional CV is useful in understanding the conformational 
dynamics of polymer chains, and represents a dense multi-variable FES which requires extensive 
visitation to resolve. The chosen system allows us to test the adaptability of 
ANN sampling to higher dimensions and compare the estimated FEL to analytical 
results for each Rouse mode $X_i$ by integrating over the others. The system 
was modeled using non-interacting beads connected via harmonic springs with bond 
coefficients of unity in reduced units. A Langevin thermostat was used to maintain 
a reduced temperature of $2/3$ integrated using the velocity-Verlet integrator 
with a timestep of $0.005$. A modified version of SSAGES v0.7.5 
compiled with LAMMPS 30Jul2016 was used for the simulation.

\begin{figure}
  \begin{center}
    \includegraphics[width=\figurewidth]{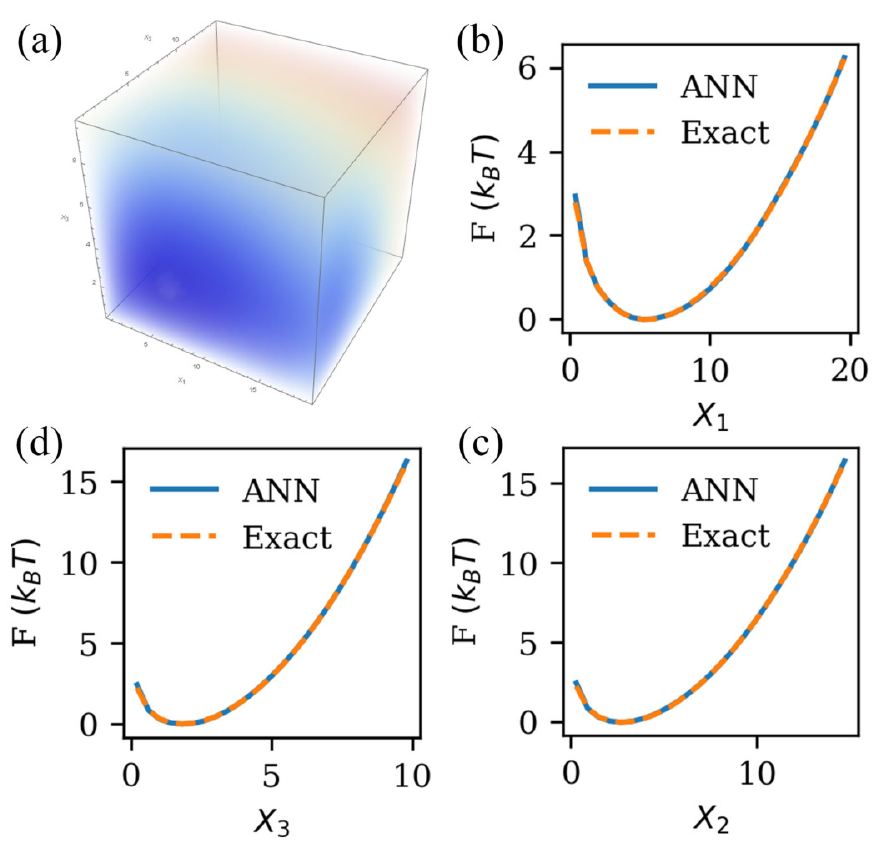}
  \end{center}
  \caption{
    Dense three dimensional Rouse mode ($X_i$) free energy landscape for a 21 bead Gaussian 
    chain. The 3D volume free energy (a) can be integrated along two chosen axes to yield 
    the one--dimensional Rouse mode free energies. Comparison with analytical theory 
    for (b) $X_1$, (c) $X_2$, and (d) $X_3$ show near exact agreement. This example demonstrates that ANN sampling should 
    be useful in sampling correlated CVs in dense high dimensional spaces. 
    }
  \label{fig:rousepanel}
\end{figure}

A $\{12, 10\}$ network with a $N = 25^3$ grid is used to converge the FEL. 
Figure~\ref{fig:rousepanel} shows a volume rendering of the final FEL and the three 
integrated Rouse modes compared to theoretical prediction. The near exact agreement 
provides an objective measure of the accuracy of ANN sampling. Additionally, the 
number of hidden layers required to represent the FEL does not necessarily increase 
with the number of dimensions. Introducing a second hidden layer when going 
from a single to two or more dimensions is necessary to efficiently represent 
multidimensional surfaces, but a further increase is not strictly required. This 
is because the level set of neuron is a hyperplane and only upon composition 
(via the addition of another hidden layer) is multidimensional nonlinearity 
naturally representable.

\begin{figure*}[t]
  \begin{center}
    \includegraphics[width=\textwidth]{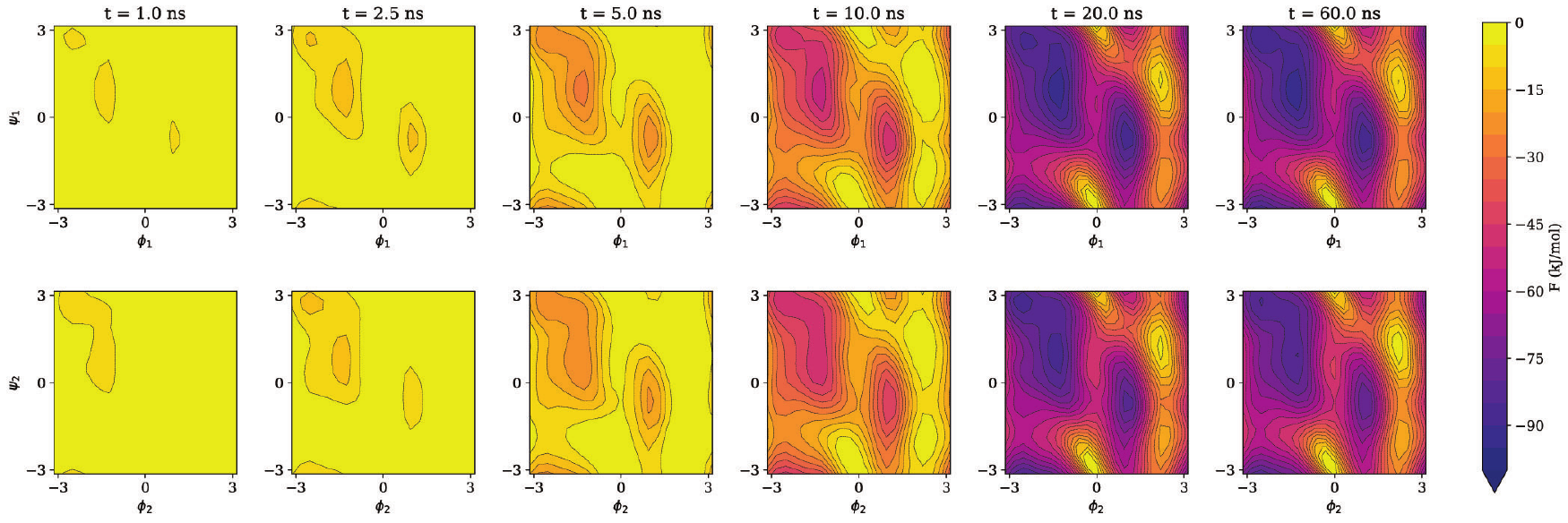}
  \end{center}
  \caption{
    Evolution of 4D free energy estimate for alanine tripeptide in vacuum shown as 
    two-dimensional Ramachandran plot projections. The top and bottom rows are the 
    first and second $\psi$-$\phi$ pairs respectively. By $5$ ns, the bias has built 
    up in the low-lying metastable configurations allowing for efficient diffusion
    of all CVs (cf. Fig.~\ref{fig:tripeptide_cvs}). The remaining time is spent resolving 
    regions of extremely low probability which converges at approximately $20$ ns. 
  }
  \label{fig:tripeptide_fes}
\end{figure*}

\subsection{Four dimensional FEL: Alanine tripeptide in vacuum}
In our final example we compute the free energy landscape of alanine tripeptide 
(Ace-Ala-Ala-Nme) in vacuum using all four backbone dihedrals as collective variables. 
This example was previously studied by Stecher et al.\cite{Stecher2014} using 
the CHARMM22 forcefield and Gaussian process regression (GPR). Here we use Amber99sb, 
hydrogen bond constraints, and a 2 fs timestep at $298.15$ K. Electrostatic interactions 
were handled using particle mesh Ewald and temperature was maintained using the
stochastic velocity rescale thermostat. As with the previous examples, a modified version 
of SSAGES v0.7.5 was used compiled with Gromacs 2016.3. 

\begin{figure}
  \begin{center}
    \includegraphics[width=\figurewidth]{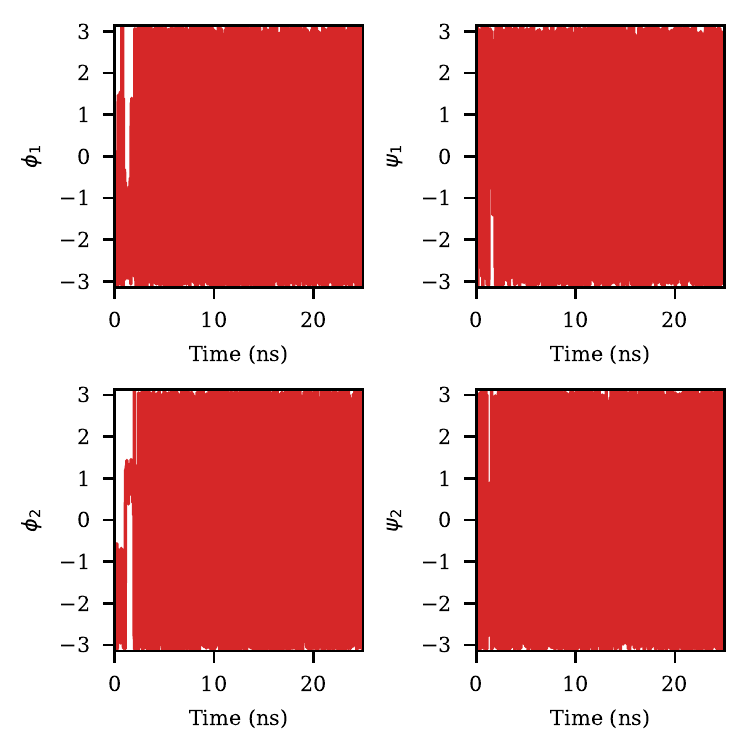}
  \end{center}
  \caption{
    Time evolution of the four backbone dihedrals of alanine tripeptide which are used 
    as collective variables with ANN sampling. $\phi_1$ and $\phi_2$ represent the slow 
    modes of the peptide which are initially confined to metastable configurations. As 
    bias accumulates each individual coordinate exhibits free diffusion across the metastable states; this occurs 
    at approximately $3$ nanoseconds.
  }
  \label{fig:tripeptide_cvs}
\end{figure}

We use a $\{24, 20\}$ network with a $N = 20^4$ grid and a sweep length of $20$ ps. As
with the previous example, there is no need to introduce a third hidden layer.
Figure~\ref{fig:tripeptide_fes} shows the evolution of the 4D free energy estimate 
projected onto two Ramachandran plots. ANN sampling rapidly builds up bias in the 
prominent free energy basins of the tripeptide, uncovering essential free energy basins and their structure within 5~ns of simulation time. This is allows the CVs to begin diffusing 
across these basins within a few nanoseconds, as shown in Figure~\ref{fig:tripeptide_cvs}. 
The remaining time is spent resolving regions of low probability, with a relative free 
energy as high as $90$ kJ/mol. Convergence is achieved at approximately $20$ nanoseconds 
with a reference FEL at a much later $60$ ns showing no appreciable change. 

In Ref.~\cite{Stecher2014}, the 4D FEL of alanine tripeptide is reconstructed using GPR 
from an increasing number of short $500$ picosecond simulation windows. Qualitative 
agreement with a reference simulation is achieved after a total aggregate simulation 
time of $1.3$ microseconds. While a direct comparison is not possible, we find that the 
$20$ nanoseconds required to resolve the 4D FEL with ANN sampling to be remarkable. 
For high dimensional systems this corresponds to a substantial improvement in
efficiency, where traditional methods can be prohibitively expensive. The results 
presented here from a single simulation, although ANN sampling is also 
trivially parallelizable. Our software implementation already contains support multiple
walkers. Similar to GPR, this would allow multiple independent simulations to contribute 
to the same FEL, taking advantage of modern parallel architectures. 

There are some additional important considerations for high dimensional CVs. 
As can be seen from our examples, we decrease the number of grid points per dimension 
as the system dimensionality increases. The simple reason for this is the curse of 
dimensionality: for a fixed point density, the training data size increases exponentially 
with increasing dimension. Furthermore, Bayesian regularization requires the calculation 
and storage of the Jacobian which scales as $\mathcal{O}(N^2)$, and requires the 
inversion of the approximate Hessian. In an effort to minimize training time, we 
investigate the effect of limiting the maximum number of training iterations on the 
performance of ANN sampling. 

The results presented above were obtained using a maximum number of $10$ iterations 
per sweep. While this seems to be unreasonably small, it performs very well as shown. 
The reason for this is that in ANN sampling, the network weights are carried over 
each training cycle. This can be seen as a form of pre-training, where the network only 
requires a minimal amount of fine-tuning to adapt to new data. We also carried out the 
simulations using $20$, $50$, and $100$ maximum iterations and found little difference 
in performance, and due to the fewer iterations incurred much less overhead. These 
limitations are a result of targeting a uniform probability distribution in the 
collective variables. Targeting a well-tempered distribution or implementing a maximum 
fill-level on the free energy can be complimented with a sparse grid, since only regions 
of CV space accessible within a desired cutoff will be accessible. This sort of approach
is often desirable and will scale better to even higher dimensions. We see this as an 
area of future research. 

\section{Discussion and Conclusions}

We have presented a novel advanced sampling algorithm utilizing
artificial neural networks to learn free energy landscapes. The
ability of ANNs to act as universal approximators enables the method
to resolve topologically complex FELs without presenting numerical
artifacts typical of other methods which rely on basis sets. Bayesian
regularization allows the entire set of training data to be used,
prevents overfitting and eliminates the need for any learning
parameters to be specified. This and other robust features result in
a method which is exceptionally swift and accurate at obtaining FELs.

As ANN sampling superficially resembles the recently proposed NN2B method~\cite{Galvelis2017}, some remarks on their similarities and differences should be made. Though it also uses neural networks to estimate free energies, it
differs significantly from out method beyond that. NN2B requires the specification
of a large number of parameters, requiring significant knowledge of CV correlation times
and sampling density, and it is not clear how they influence convergence. Our use of a 
discretized global partition function estimate retains information on all CV states 
sampled over the course of a simulation, while NN2B requires a judicious choice of 
the degree of subsampling from prior sweeps. The use of standard neural networks 
and their density estimation procedure necessitates significantly larger architectures 
and sampling intervals to represent a given free energy landscape. It also means 
that some of the valuable data must be used for validation to avoid overfitting.  

The consequences of this are reflected in their study of alanine dipeptide in vacuum. 
Training times for each sweep are reported to range from 10 to 30 minutes depending 
on the size of the network. Furthermore, it requires 50 nanoseconds to converge the 
two dimensional FEL for ADP in vacuum. This is considerably longer than the typical 
2 ns using standard metadynamics. In comparison, our method requires seconds 
to train the 2D FEL and a minute for higher dimensions in the early stages; as the 
approximate FES becomes smoother, the training time decreases considerably due to 
regularization. Both the lengthy training  times and slow convergence render NN2B 
impractical for most applications. We do however see merit in using nearest neighbor 
density estimators in place of sparse histograms, and anticipate that this approach 
can be integrated within our proposed method. 

The considerable advantages we have demonstrated biasing with ANN networks position this 
as an ideal method for the study of computationally expensive systems where sampling is 
limited. In particular, both first--principles MD and large biomolecule simulations 
are prime candidates for ANN sampling,
where all but the most trivial of examples become intractable due to
the typical time scales required to obtain free energy estimates. The
significant speed improvements in sampling afforded by this method
also open the door to the use of accelerated free-energy calculations
for high throughput computational screening of material properties
(see, e.g.~\onlinecite{Wilmer2012}). The ANN framework also makes it
possible to initialize simulations with pre--trained networks, a
technique commonly used in deep learning applications. Starting with a
FEL from a classical simulation can considerably improve the
convergence of a first--principles estimate, while a theoretical or
coarse--grained model can inform the initial bias in a biomolecular
simulation. ANN sampling is a promising new approach to advanced
sampling which can help resolve complex free energy landscapes in less
time than conventional approaches while overcoming many previous
shortcomings.

\section{Data availability}

All of the run files, data, and analysis scripts required to reproduce the 
results in this paper are freely accessible online at 
\url{https://github.com/hsidky/ann_sampling}. ANN sampling will be available 
in the next SSAGES release.

\section{Acknowledgements}

HS acknowledges support from the National Science Foundation Graduate
Fellowship Program (NSF-GRFP). HS and JKW acknowledge computational
resources at the Notre Dame Center for Research Computing (CRC). HS and JKW
acknowledge Michael Webb (U. Chicago) for providing the analytical Rouse mode solutions. This
project was supported by MICCoM, as part of the Computational
Materials Sciences Program funded by the U.S. Department of Energy,
Office of Science, Basic Energy Sciences, Materials Sciences and
Engineering Division.



%
  
\end{document}